\documentclass[
]{ceurart}

\usepackage{listings}
\usepackage{subfig}
\usepackage{caption}
\usepackage{subcaption}
\begin{document}

\copyrightyear{2024}
\copyrightclause{Copyright for this paper by its authors.
  Use permitted under Creative Commons License Attribution 4.0
  International (CC BY 4.0).}

\conference{CHIRP 2024: Transforming HCI Research in the Philippines Workshop, May 09, 2024, Binan, Laguna}

\title{Point n Move: Designing a Glove-Based Pointing Device}

\author[]{Sealtiel B. Dy}[%
email=sealtiel_dy@dlsu.edu.ph,%
]
\author[]{Robert Joachim O. {Encinas}}[%
email=robert_joachim_encinas@dlsu.edu.ph,%
]
\author[]{Daphne Janelyn L. {Go}}[%
email=daphne_janelyn_go@dlsu.edu.ph,%
]
\author[]{Kyle Carlo C. {Lasala}}[%
email=kyle_lasala@dlsu.edu.ph,%
]
\author[]{Bentley Andrew Y. {Lu}}[%
email=bentley_andrew_lu@dlsu.edu.ph,%
]
\author[]{Maria Monica {Manlises}}[%
email=maria_monica_manlises@dlsu.edu.ph,%
]
\author[]{Jordan Aiko {Deja}}[%
email=jordan.deja@dlsu.edu.ph,%
orcid=0001-9341-6088%
]

\address[1]{De La Salle University, Manila, Philippines}


\begin{abstract}
In-person presentations commonly depend on projectors or screens, requiring input devices for slide transitions and laser pointing. This paper introduces a glove-based pointer device that integrates these functions, offering an alternative to conventional tools. The device leverages accelerometer and gyroscope technology to enhance precision and usability. We evaluated its performance by comparing it to the original CheerPod interface in hierarchical menu navigation tasks, involving participants aged 18 to 25. Results indicate task completion times ranging from 9 to 15 seconds with the proposed device, highlighting its efficiency and consistency. While the original CheerPod interface performed adequately, the glove-based pointer demonstrated advantages in reliability across tasks. These findings contribute to the design considerations for wearable input devices and suggest pathways for future improvements in presentation tools.
\end{abstract}


\maketitle

\section{Introduction}

\begin{figure}
 \includegraphics[width=\textwidth]{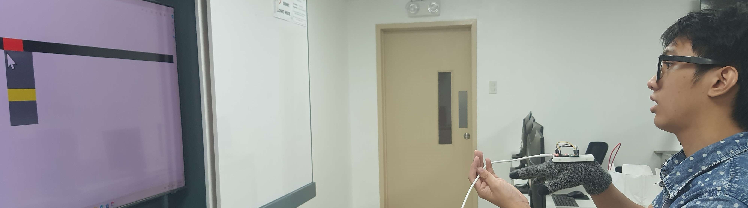}
  \caption{Preview: A participant performing a navigation task using our ``Point n Move'' interface. It involves the use of a modified glove-based pointing device.}
  \label{fig:teaser}
\end{figure}

\par Presenters often use projectors connected to laptops or desktop computers for in-person presentations, requiring multifunctional devices to streamline presentation flow and enhance control. Key functionalities include remote slide navigation, section highlighting, and mouse control. CheerDots developed the CheerPod to address these needs, integrating mouse and laser pointer capabilities. Measuring 2.6 inches long and weighing 33 grams, the CheerPod connects via Bluetooth to personal computers, smartphones, and tablets \cite{cheerdotsBuyPage}. It operates in two modes: ground mode, functioning as a mouse for hovering, clicking, scrolling, and swiping, and air mode, acting as a remote touchpad and laser pointer for zooming, clicking, and slide navigation \cite{cheerdots}. This study builds on the CheerPod’s design, focusing on improving its navigation capabilities for projected screens. The proposed design aims to offer an alternative pointer device specifically for presentation use, retaining all key features of the original while addressing identified limitations. Other use cases will be observed but are not the focus of this research. To evaluate the device, we define four main research questions: Does the proposed pointer device retain the key features of the original? (\textbf{RQ 1}). Is the proposed pointer device reliable and usable for intended use cases? (\textbf{RQ 2}). Does the proposed pointer device match or exceed the performance of the original? (\textbf{RQ 3}). Is the proposed pointer device intuitive and user-friendly? (\textbf{RQ 4}).

\par Key limitations of the CheerPod identified as a basis for improvement include restricted cursor movement, limited concurrent actions, and a disconnect between laser pointing and cursor movement. These challenges stem from the CheerPod’s handheld design, which the proposed device seeks to address through a redesigned form factor. 

\par This paper presents the design, development, and evaluation of a glove-based pointer device as an alternative to existing presentation tools like the CheerPod. The proposed device integrates an accelerometer and gyroscope sensor with a microcontroller to enable intuitive hand movements for controlling pointer functions in presentation tasks. The study compares the performance of two iterations of the prototype against the CheerPod, focusing on metrics such as task completion time, stability, and user intuitiveness. Through this comparison, the paper highlights the strengths and limitations of the proposed prototype and suggests areas for future improvements in hardware and functionality, offering valuable insights into the design of more effective presentation tools.

\section{CheerPod: Redesigned Interface}

\par The proposed device is a microcontroller-based prototype designed to enhance the functionality of presentation tools by leveraging a sensor as its input device. Attached to a glove, the prototype uses hand movements as its primary input, allowing presenters the flexibility to use both hands for props or other actions during presentations. The complete system consists of two key components: (1) an accelerometer and gyroscope sensor and (2) a microcontroller unit. These components work together to process and interpret motion data. The design and functionality of the prototype draw inspiration from earlier work, including \cite{abdelkader_ch_2021} and \cite{neuromodulator_2021}, which informed its development.


\subsection{Hardware Used}

\begin{minipage}{\linewidth}
    \centering
    \includegraphics[width=\linewidth]{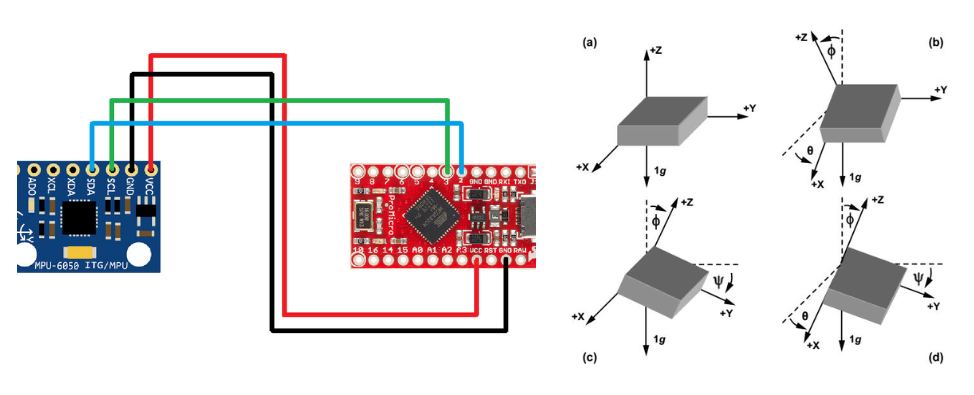}
    \captionof{figure}{Schematic Diagram of Circuit Implementation Angle of Inclination of Accelerometer in MPU 6050}
    \label{inputCircuit}
\end{minipage}

\par The prototype consists of an Arduino Pro Micro, an MPU-6050 sensor, breadboards, and male-to-male cables, all connected as shown in the circuit schematic in Figure \ref{inputCircuit}.

\par Initially, an Arduino Nano was selected due to its compact size \cite{arduinoNano}. However, the need for additional Python integration to transmit sensor data introduced complexity and delays. To address this, the Arduino Pro Micro was chosen instead, leveraging its built-in USB communication to streamline data processing and facilitate mouse movement portability \cite{arduinoMicro, buxto_1983}.

\par The MPU-6050 sensor serves as the motion-tracking component, combining a 3-axis gyroscope and a 3-axis accelerometer to detect motion, orientation, and acceleration across a 3D plane. It processes changes in position, tilt, inclination, or movement along the X, Y, and Z axes, as depicted in Figure \ref{inputCircuit} \cite{electronicwings_2020}. The embedded Digital Motion Processor (DMP) computes motion data, including roll, pitch, yaw angles, and orientation, reducing the need for additional processing. However, after two minutes of continuous use, the sensor experiences instability, leading to drift and decreased pointer accuracy \cite{electronicwings_2020}. This limitation arises from the lack of a magnetometer for absolute orientation. A reset after two minutes is recommended to mitigate these issues. Future iterations may incorporate sensors like the MPU-9250 or an external magnetometer such as the HMC5883L to enhance stability and accuracy.

\par The firmware, implemented in Arduino, processes motion data from the MPU-6050 to control the mouse cursor on a connected device. The I2C Device Library simplifies communication between devices via the I2C protocol \cite{rowberg_2022}. This library provides built-in functions for reading, writing, and configuring data. The prototype uses these functions to process raw motion data, applying a transfer function to translate movement into cursor motion relative to screen dimensions. Before any operation, the DMP is calibrated to zero using a buffer that precedes sensor data reading. The yaw, pitch, and roll values are then computed and displayed on a serial monitor. Different transfer functions with varying thresholds and sensitivities were applied in two iterations of the prototype \cite{rowberg_2022}.

\par The final prototype integrates a glove, allowing users to operate the device while keeping one hand free for other tasks. As shown in Figure \ref{fig:glove}, the circuit board is mounted on the glove, and users must position their hand in a specific orientation to control the pointer on the screen. Similar to trackball transducers, the glove supports only motion tracking (State 1) and lacks a button-down state (State 2) \cite{buxton_1990}. This design ensures flexibility and usability during presentations.

\begin{minipage}{\linewidth}
    \centering
    \includegraphics[width=0.3\linewidth]{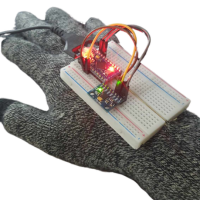}
    \captionof{figure}{Final Glove-Based Prototype}
    \label{fig:glove}
  \end{minipage}

\section{Method}
\par To evaluate the proposed pointer device in the context of public presentations, the experimental setup was designed to closely replicate a typical presentation environment. A large projected display was used for testing, as large displays have been shown to support high touch performance despite some limitations, such as fatigue during prolonged use \cite{listkipp_2019}. These limitations were deemed negligible given the short-duration use typical in presentation scenarios.

\par The primary quantitative metric for evaluating device effectiveness was the average movement time (MT) required to complete the given navigation tasks. This metric aligns with the effectiveness measures outlined in \cite{card_mackinlay_robertson_1990}. The Accot-Zhai Steering Law, specifically the average MT, was used as the framework for comparison, as the navigation task involved pointer movement constrained by strict boundaries \cite{accot_zhai_1997}. Additionally, the devices were evaluated on time-to-grasp, another metric from \cite{card_mackinlay_robertson_1990}, which was quantified as the slope of the MT trendline over trial runs.

\par The navigation task was adapted from the hierarchical menu example in \cite{laubheimer_2019} and implemented as an HTML page that simulates an abstract version of the example, as shown in Figure \ref{fig:basis}. This setup allowed for a standardized and controlled evaluation of the pointer devices.



\begin{minipage}{\linewidth}
    \centering
    \includegraphics[width=\linewidth]{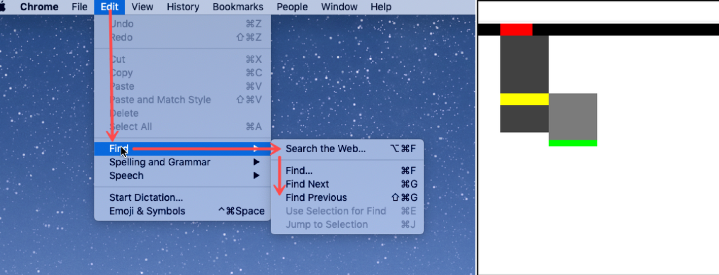}
    \captionof{figure}{Navigation Task from \cite{laubheimer_2019}}
    \label{fig:basis}
\end{minipage}

\par The navigation task required participants to hover the cursor over highlighted buttons in the interface. The HTML page was displayed on a large TV screen, and participants positioned themselves at a comfortable distance, up to a maximum of 3 meters from the screen. Each experiment consisted of four trials, with each trial involving four randomly placed targets. The movement times for all targets in each trial were recorded. This setup was replicated for three groups: one using the CheerPod, one using iteration 1 of the prototype, and one using iteration 2 of the prototype.

\par A total of 24 participants, aged 18 to 25, participated in the study, evenly distributed across the three groups. This age group was selected due to its familiarity with digital interfaces, which is relevant for tasks involving technological devices.

\par A comparative analysis was performed on the movement times across the four trials for the three groups: the control group (CheerPod), the prototype iteration 1 (low sensitivity with high X-Y threshold values), and the prototype iteration 2 (high sensitivity with low X-Y threshold values). A T-test was conducted to assess whether the differences in performance among the three devices were statistically significant. These results helped determine if the proposed prototypes demonstrated measurable improvements over the original interface.

\par In addition to statistical analysis, performance metrics were visually compared across the three groups using data visualization techniques, including box plots, histograms, and line graphs. These visualizations provided a detailed understanding of the performance differences among the CheerPod and the two prototype iterations.

\section{Results}

  \begin{figure}
      \centering
      \includegraphics[width=1\linewidth]{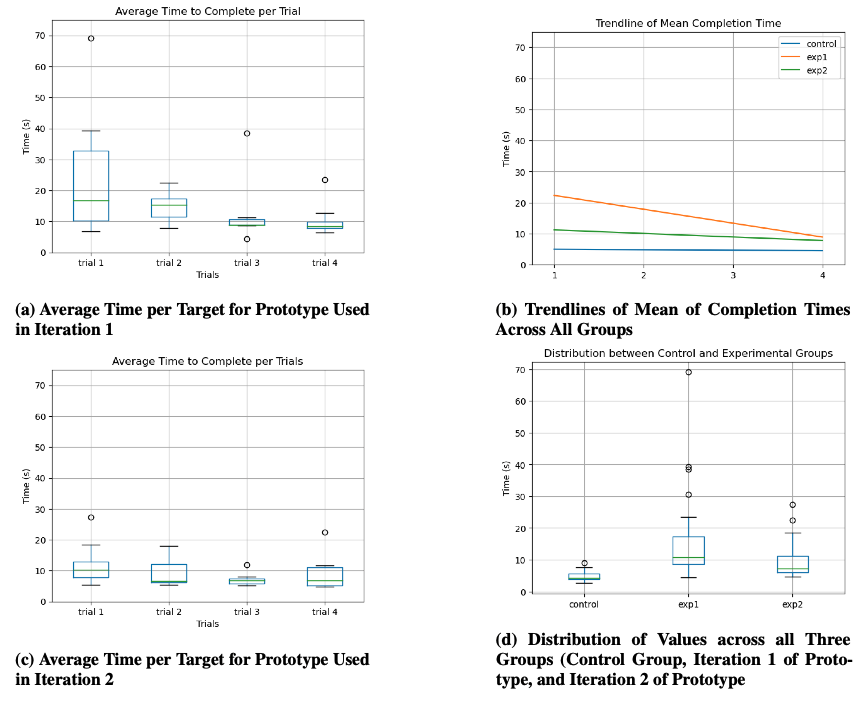}
      \caption{Various figures showing different plots of times taken when using the two iterations of the prototype.}
      \label{fig:datacharts}
  \end{figure}

\par Figure \ref{fig:datacharts}(a) presents the average time taken by eight participants to complete each of the eight trials using prototype iteration 1, which employed low sensitivity and high X-Y threshold values. The results indicate that Trial 1 took noticeably longer on average compared to subsequent trials, suggesting a potential learning curve or adaptation period for the participants.

\par Figure \ref{fig:datacharts}(c) illustrates the average time taken by eight participants to complete the same trials using prototype iteration 2, which featured high sensitivity and low X-Y threshold values. In this case, the task completion times were consistent across all trials, indicating improved ease of use or adaptability with this iteration.

\par Figure \ref{fig:datacharts}(d) compares the distribution of completion times between the control group (using the CheerPod) and the experimental groups (iterations 1 and 2 of the prototype). The results show that participants using the CheerPod achieved the lowest average completion times and exhibited the least variability in performance, outperforming both prototype iterations.

\par These findings suggest that while both prototypes show promise, the CheerPod remains more reliable and efficient under the conditions tested, though the second iteration of the prototype demonstrates greater consistency compared to the first.



\subsection{Insights About the Proposed Device}
\par The evaluation of the proposed pointer device demonstrated that participants completed tasks within 9 to 15 seconds across both iterations, confirming the device's basic usability and reliability. However, the CheerPod significantly outperformed both iterations, with an average completion time of 4.75 seconds and a p-value < 0.000 compared to the prototypes (see Tables \ref{tab:control_vs_day1} and \ref{tab:control_vs_day2}). This discrepancy may stem from challenges in performing diagonal movements with the prototype, as discussed in \cite{guiard1987asymmetric}.

\par Figure \ref{fig:datacharts}(b) highlights further differences: the control group (CheerPod) exhibited a nearly flat trendline across trials ($slope = -0.0159$), indicating stable performance, whereas the experimental groups showed steeper slopes of $-4.474$ in Iteration 1 and $-1.144$ in Iteration 2. This suggests that the CheerPod offered greater stability and a smoother learning curve, likely due to users' familiarity with its touchpad interface.

\par Finally, a p-value of 0.017 between the first and second prototype iterations (see Table \ref{tab:day1_vs_day2}) indicates a statistically significant performance improvement in Iteration 2, reflecting the benefits of design adjustments such as higher sensitivity and lower X-Y thresholds. These results underline the need for further refinements to enhance the usability and efficiency of the proposed device while addressing the limitations observed.



\subsection{Proposed Device Across All Iterations}

\par When comparing Iteration 1 and Iteration 2 of the proposed prototype, a significant improvement was observed in Iteration 2, as reflected by a smoother learning curve (slope improved from $-4.474$ to $-1.144$). This improvement is likely due to the transition from low to high sensitivity and adjustments to the X-Y threshold values. Despite these enhancements, the CheerPod consistently outperformed both prototype iterations, demonstrating superior performance, stability, and user-friendliness. These results underscore the effectiveness of the CheerPod while highlighting areas for further refinement in the proposed design to close the performance gap.


\subsection{Recommendations on Improving the Design of the Proposed Device}
\par While this study primarily employs a quantitative approach, incorporating a qualitative assessment could provide a more comprehensive evaluation of the prototype's performance. Additionally, extending the analysis to include the index of difficulty, as defined in the Accot-Zhai Steering Law~\cite{accot_zhai_1997, mackenzie1992extending, fitts1954information}, alongside the average movement time, could offer further insights into device performance.

\par Hardware improvements could also enhance the prototype's overall effectiveness. For example, exploring the use of the MPU-9250 accelerometer and gyroscope sensors, which include magnetometers, could help reduce tracking drift, minimize the need for frequent recalibration, and provide better independence from device orientation.

\par Moreover, integrating a Bluetooth module would enable wireless operation, addressing limitations imposed by a wired connection. To further improve functionality, a clicking feature could be added using tact switches. These switches should be strategically placed to allow clicking with one hand, thereby enabling concurrent actions and facilitating the use of State 2 \cite{buxton_1990}.

\par As the prototype evolves, it is crucial to retain the core features of the original device, as these form the minimum set of functionalities necessary for effective performance. Future iterations should aim to build on these features while introducing new enhancements.



\section{Conclusion}
\par The prototypes, in both iteration 1 and 2, maintained key features similar to the CheerPod, as outlined in Section 1.2, necessary for completing the navigation task (\textbf{RQ1}). While the first iteration showed slower and less stable results, the second iteration showed significant improvements in both stability and completion time, due to adjustments in sensitivity and X-Y threshold values (\textbf{RQ2}). However, the CheerPod outperformed the prototype in terms of both stability and faster completion time, highlighting the hardware limitations of the proposed design (\textbf{RQ3}). Additionally, participants found the prototype less intuitive than the CheerPod, as it lacked support for diagonal movements and relied on a wired interface, which affected usability (\textbf{RQ4}).

\bibliography{main}

\newpage
\appendix

\section*{Appendix: Additional Results Tables}
\hfill
\begin{table}[h]
\centering
\caption{Comparison between Control Group and Iteration 1 of Experimental Group}
\label{tab:control_vs_day1}
\begin{tabular}{@{}lll@{}}
\toprule
\textbf{Assumption}              & \textbf{t-value} & \textbf{p-value} \\ \midrule
Equality of Population Variance Assumed & -4.687           & $< 0.000$        \\
Equality of Population Variance Not Assumed & -4.687           & $< 0.000$        \\ \bottomrule
{\footnotesize{*Significant at 0.05 level of significance}}
\end{tabular}
\end{table}
\hfill

\begin{table}[h]
\centering
\caption{Comparison between Control Group and Iteration 2 of Experimental Group}
\label{tab:control_vs_day2}
\begin{tabular}{@{}lll@{}}
\toprule
\textbf{Assumption}              & \textbf{t-value} & \textbf{p-value} \\ \midrule
Equality of Population Variance Assumed & -4.811           & $< 0.000$        \\
Equality of Population Variance Not Assumed & -4.811           & $< 0.000$        \\ \bottomrule
{\footnotesize{*Significant at 0.05 level of significance}}
\end{tabular}
\end{table}
\hfill

\begin{table}[h]
\centering
\caption{Comparison between Iteration 1 and Iteration 2}
\label{tab:day1_vs_day2}
\begin{tabular}{@{}lll@{}}
\toprule
\textbf{Assumption}              & \textbf{t-value} & \textbf{p-value} \\ \midrule
Equality of Population Variance Assumed &  2.452          & $ 0.017$        \\
Equality of Population Variance Not Assumed & 2.452           & $ 0.019$        \\ \bottomrule
{\footnotesize{*Significant at 0.05 level of significance}}
\end{tabular}
\end{table}
\hfill
\begin{table}[h]
    \centering
    \caption{Average Completion Time per Group}
    \label{tab:descriptive}
    \begin{tabular}{@{}lll@{}}
    \toprule
    \textbf{Group}              & \textbf{Mean} & \textbf{Standard Deviation} \\ \midrule
    Control Group &  $04.75$& $01.42$\\
    Experimental Group Iteration 1 & $15.62$           & $13.04$        \\
    Experimental Group Iteration 2 & $09.50$& $05.40$\\
    \bottomrule
\end{tabular}
\end{table}

\end{document}